\documentstyle[11pt]{article}

\def\dalemb#1#2{{\vbox{\hrule height .#2pt
        \hbox{\vrule width.#2pt height#1pt \kern#1pt
                \vrule width.#2pt}
        \hrule height.#2pt}}}

\let\a=\alpha \let\b=\beta

\def\nn{\nonumber} \def\bd{\begin{document}} \def\ed{\end{document}}
\def\ds{\documentstyle} \let\fr=\frac \let\bl=\bigl \let\br=\bigr
\let\Br=\Bigr \let\Bl=\Bigl 
\let\bm=\bibitem
\let\na=\nabla
\let\pa=\partial \let\ov=\overline 
\newcommand{\be}{\begin{equation}} 
\newcommand{\ee}{\end{equation}} 
\def\ba{\begin{array}}
\def\ea{\end{array}}
\def\ft#1#2{{\textstyle{{\scriptstyle #1}\over {\scriptstyle #2}}}}
\def\fft#1#2{{#1 \over #2}}
\def\del{\partial}
\def\sst#1{{\scriptscriptstyle #1}}
\def\oneone{\rlap 1\mkern4mu{\rm l}}
\def\ie{{\it i.e.\ }}
\def\via{{\it via}}
\newcommand{\ho}[1]{$\, ^{#1}$}
\newcommand{\hoch}[1]{$\, ^{#1}$}
\newcommand{\bea}{\begin{eqnarray}} 
\newcommand{\eea}{\end{eqnarray}} 
\newcommand{\ra}{\rightarrow}
\newcommand{\lra}{\longrightarrow}
\newcommand{\Lra}{\Leftrightarrow}
\newcommand{\ap}{\alpha^\prime}
\newcommand{\bp}{\tilde \beta^\prime}
\newcommand{\tr}{{\rm tr} }
\newcommand{\Tr}{{\rm Tr} } 
\newcommand{\NP}{Nucl. Phys. }
\newcommand{\tamphys}{\it Center for Theoretical Physics,
Texas A\&M University, College Station, Texas 77843}
\newcommand{\ens}{\it Laboratoire de Physique Th\'eorique de l'\'Ecole
Normale Sup\'erieure\hoch{2}\\
24 Rue Lhomond - 75231 Paris CEDEX 05}

\newcommand{\auth}{H. L\"u\hoch{\dagger} and
C.N. Pope\hoch{\ddagger1}}

\begin{document}
\begin{flushright}
\hfill{CTP TAMU-6/97}\\
\hfill{LPTENS-97/03}\\
\hfill{hep-th/9701177}\\
\hfill{Jan 1997}\\
\end{flushright}

\vspace{15pt}

\begin{center}
{ \large {\bf T-duality and U-duality in toroidally-compactified strings}}

\vspace{20pt}
\auth

\vspace{15pt}

{\hoch{\dagger}\ens}

\vspace{10pt}
{\hoch{\ddagger}\tamphys}

\vspace{40pt}

\underline{ABSTRACT}
\end{center}

     We address the issue of T-duality and U-duality symmetries in the
toroidally-compact- ified type IIA string.  It is customary to take as
a starting point the dimensionally-reduced maximal supergravity
theories, with certain field strengths dualised such that the
classical theory exhibits a global $E_{n(n)}$ symmetry, where $n=11-D$
in $D$ dimensions.  A discrete subgroup then becomes the conjectured
U-duality group.  In dimensions $D\le 6$, these necessary dualisations
include NS-NS fields, whose potentials, rather than merely their field
strengths, appear explicitly in the couplings to the string
worldsheet.  Thus the usually-stated U-duality symmetries act
non-locally on the fundamental fields of perturbative string theory.
At least at the perturbative level, it seems to be more appropriate to
consider the symmetries of the versions of the lower-dimensional
supergravities in which no dualisations of NS-NS fields are required,
although dualisations of the R-R fields are permissible since these
couple to the string through their field strengths.  Taking this
viewpoint, the usual T-duality groups survive unscathed, as one would
hope since T-duality is a perturbative symmetry, but the U-duality
groups are modified in $D\le 6$.

{\vfill\leftline{}\vfill
\vskip	10pt
\footnoterule
{\footnotesize	\hoch{1} Research supported in part by DOE 
Grant DE-FG05-91-ER40633 \vskip	-12pt} \vskip 10pt
{\footnotesize
        \hoch{2} Unit\'e Propre du Centre National de la Recherche
Scientifique, associ\'ee \`a l'\'Ecole Normale Sup\'erieure et \`a
l'Universit\'e de Paris-Sud}} 

\pagebreak
\setcounter{page}{1}

\bigskip\bigskip\bigskip\bigskip

    Eleven-dimensional supergravity has enjoyed a chequered history
since its discovery in 1978 \cite{cjs}.  Since it occupies the
distinguished position of being the highest-dimensional supergravity
theory, it has long been thought likely to play an important r\^ole in
fundamental physics.  However, earlier attempts to exploit it as a
starting-point for a unified theory foundered for a variety of reasons,
including its non-renormalisability and the difficulties of extracting a
realistic-looking four-dimensional theory from it.  Nonetheless, the
developments that were made in the process of trying to achieve a
realistic theory established many of the ideas that
have subsequently been used extensively in string theory.  

    One of the striking features of $D=11$ supergravity is that its
dimensional reduction on a circle gives rise, after truncation to the
massless sector, to precisely the low-energy effective limit of the
ten-dimensional type IIA string.  The true significance of this fact
really only emerged with the observation by Witten that the
$D=11$ theory can be viewed as describing the degrees of freedom of the
type IIA string in the limiting regime where its coupling constant
tends to infinity \cite{w1}.  This can be seen in the Kaluza-Klein
description of the reduction from $D=11$ to $D=10$, in which the
Einstein-frame metrics of the two theories are related by
\be
ds_{11}^2 = e^{\ft16 \phi}\, ds_{\sst E}^2 + e^{-\ft43 \phi}\, (dz+ {\cal
A} )^2\ ,\label{ms}
\ee
where $\phi$ is the dilaton of the IIA string, $z$ is the eleventh
coordinate which is compactified on a circle, and ${\cal A} = {\cal
A}_{\sst M}\, dx^{\sst M}$ is the Ramond-Ramond vector potential of
the IIA string.  Thus we see that in the limit where the string
coupling constant $g=e^{-\phi_0}$ becomes large, the radius $R_{11}
=e^{-2/3\phi_0}$ of the eleventh dimension enlarges such that the
description of the IIA string theory effectively becomes eleven
dimensional.  On the other hand, the eleventh dimension becomes invisible
in perturbative string theory, where the coupling $g$ is small.
Of course it is no longer believed that supergravity itself is the
fundamental theory in $D=11$; rather, there is an as-yet undiscovered
M-theory whose low-energy limit is described by $D=11$ supergravity.

      The M-theory conjecture would be further strengthened if one
found also that the physical degrees of freedom in $D=10$ had their
origin in $D=11$.  In particular, massless states in $D=11$ will give
rise not only to massless states but also to massive states 
upon circular compactification to $D=10$, where the masses of the
particles are given by $n/R_{11}$.  Indeed, there do exist extremal black
hole solutions carrying R-R electric charges in type IIA supergravity,
which correspond to Kaluza-Klein charges from the eleven-dimensional
point of view.  In order to discuss the masses of these black holes,
for comparison with the massive Kaluza-Klein states of M-theory, it
is necessary first to specify the metric with respect to which the
masses are being measured.  The metric $ds_{\sst E}^2$ in (\ref{ms})
is the Einstein-frame metric in $D=10$, which we shall denote by
$G_E$.  In ten-dimensional string theory, there are two more metrics
that are relevant: one is the string metric $G_S$, related to the
eleven-dimensional metric by $ds_{11}^2 = e^{\ft23 \phi} ds_{\sst S}^2
+ e^{-\ft43\phi} (dz+ {\cal A})^2$, and the other is the same as the
eleven-dimensional metric $G_{M}$ without any conformal rescaling,
{\it i.e.}\ $ds_{11}^2 = ds_{\sst M}^2 + e^{-\ft43\phi} (dz+ {\cal
A})^2 $. Thus we have 
\be
G_{M} = e^{\ft23\phi}\, G_S = e^{\ft16\phi}\, G_E\ .\label{metricrl}
\ee
For a generic extremal $p$-brane, if the mass per unit $p$-volume is $m_A$
in the metric $G_A$ and $m_B$ in the metric $G_B$, and if the two metrics
are conformally related by $G_A= \Omega^2 G_B$, then from simple
dimensional analysis we have
\be
m_{\sst B} = \Omega_0^{p+1} m_{\sst A}\ ,\label{massrl}
\ee
where $\Omega_0$ is the asymptotic value of $\Omega$ at infinity.  The
black hole solutions in the $D=10$ type IIA theory have masses given
by $e^{\ft34\phi_0}Q$ in the Einstein-frame metric \cite{dkl}, where
$Q$ denotes the electric charge.  It follows from (\ref{metricrl}) and
(\ref{massrl}) that the masses are $Q/g$ in the string metric $G_S$,
and $Q/R_{11}$ in the eleven-dimensional metric \cite{w1}.  Thus from
the string-metric point of view, the black hole solutions are
intrinsically non-perturbative, and thus are absent from the
perturbative string spectrum.  From the eleven-dimensional point of
view, the masses are indeed precisely the same as those of the
Kaluza-Klein massive modes coming from M-theory.

         There are also other extremal $p$-brane solutions in $D=10$
type IIA supergravity for other values of $p$, whose masses in the
Einstein-frame metric can easily be obtained \cite{dkl}.  To be
precise, if the coupling of a field strength to the dilaton is given
by $e^{-a\phi} F^2$ in the Einstein-frame metric, then the mass per
unit $p$-volume for the corresponding electrically-charged solution is
given by $e^{\ft12a\phi_0}Q$, whilst for the magnetically-charged
solution it is given by $e^{-\ft12a\phi_0}Q$.  From these masses, we
can enumerate the masses in the string metric $G_S$ and in the
eleven-dimensional metric $G_M$ for all the $p$-branes in type IIA
supergravity.  They are summarised in table 1:

\vfill\eject

\bigskip\bigskip

\centerline{
\begin{tabular}{|c|c|c|c|c|c|c|c|}\hline
\multicolumn{2}{|c|}{$D=11$} & \multicolumn{2}{c|}{membrane}
& \multicolumn{2}{c|}{5-brane} & pp-wave
& twisted \\ \hline
\multicolumn{2}{|c|}{$D=11$ mass} & \multicolumn{2}{c|}{$Q_2$} &
\multicolumn{2}{c|}{$Q_5$} &  $Q_{pp}$ & $Q_t$ \\ \hline
\multicolumn{2}{|c|}{$D=10$} & string & 2-brane & 4-brane &
5-brane & 0-brane & 6-brane \\ \hline
$D=10$ mass& $G_{M}$ & $R_{11} Q_2$ & $Q_2$ & $R_{11} Q_5$ & $Q_5$ & 
$R_{11}^{-1} Q_{pp}$ & $R_{11}^2 Q_t$ \\ \hline
$D=10$ mass& $G_S$ & $Q_2$ & $g^{-1} Q_2$ & $g^{-1} Q_5$ & $g^{-2} Q_5$ &
$g^{-1} Q_{pp}$ & $g^{-1} Q_t$ \\ \hline
$D=10$ mass& $G_E$ & $e^{-\ft12\phi_0} Q_2$ & $e^{\ft14\phi_0} Q_2$ &
$e^{-\ft14\phi_0} Q_5$ & $e^{\ft12 \phi_0} Q_5$ &
$e^{\ft34\phi_0} Q_{pp}$ & $e^{-\ft34\phi_0} Q_t$ \\ \hline   
\end{tabular}}

\bigskip

\centerline{Table 1:  $p$-brane masses in the various $D=11$ and $D=10$
metrics}
\bigskip\bigskip

     Note that although the string tension vanishes in both the
Einstein-frame metric $G_E$ and the eleven-dimensional metric $G_M$
when the radius $R_{11}$, and hence the coupling constant $g$, become
small, it remains a constant in the string metric $G_S$.  (In fact,
all the $p$-brane tensions are independent of the dilaton coupling in
their own metric.) In the string metric, for all the $p$-branes with
R-R charges, and the 5-brane, the mass per unit $p$-volume goes to
infinity when $g$ becomes small, implying that they describe
non-perturbative degrees of freedom.  The masses of the
ten-dimensional string and membrane in the eleven-dimensional metric
$G_M$ are $R_{11} Q_2$ and $Q_2$, consistent with the fact that they
are obtained from double \cite{dhis} and vertical \cite{t1}
dimensional reduction of the eleven-dimensional membrane.  The same
analysis applies to the ten-dimensional 4-brane and 5-brane.  All the
$p$-branes in type IIA supergravity describe physical degrees of
freedom of the string theory, since they are BPS saturated states
which will survive quantisation. As we saw, these $p$-branes all have
natural eleven-dimensional explanations.  This implies that the
membranes, 5-branes, {\it etc.} in $D=11$ are all part of the physical
degrees of freedom of M-theory.

     The above discussion relates the type IIA string in ten dimensions with
an eleven-dimensional theory.  Previous work, notably that by Hull and
Townsend, had already provided strong indications that $D=11$
supergravity has a r\^ole to play in describing the properties of type
IIA or IIB strings compactified to $D\le 9$ dimensions on a torus
\cite{ht}.  In particular, it was conjectured that the
long-established Cremmer-Julia global symmetry group $E_{n(n)}$ of
maximal supergravity in $D=11-n$ dimensions \cite{cj1,cj2} survives in
the associated toroidally-compactified string theories in the form of
a discrete subgroup $E_{n(n)}(Z)$ that is an exact U-duality symmetry
at the full quantum non-perturbative level \cite{ht}.  This U-duality
symmetry is the closure of the $D_{n-1}\sim O(n-1,n-1)$ subgroup of
perturbative T-duality symmetries and an $SL(2,R)$ subgroup which 
describes a non-perturbative duality symmetry that interchanges the
NS-NS and R-R fields of the theory.  By contrast the T-duality
symmetry, which is valid order-by-order in string perturbation theory,
acts only within the NS-NS and R-R sectors themselves, but does not
mix between them.

    One of the issues arising in the discussion of U-duality symmetries 
in compactified string theories is the question of whether these
symmetries are simply artefacts of the compactification procedure, or
whether they are providing insights into the symmetries of the
ten-dimensional strings themselves.  In fact the answer seems to lie
somewhere in between.  One illustration of this is provided by
considering the example of the low-energy limit of the type IIA
string.  In ten dimensions this supergravity theory has no
Cremmer-Julia type symmetry group; however, upon compactification on a
circle to $D=9$ one finds an $SL(2,R)$ global symmetry (actually
$GL(2,R)$ \cite{cj2}, including an additional $R$ symmetry that is not
relevant for our immediate discussion).  Now $SL(2,R)$ is a symmetry
that is characteristic of the Kaluza-Klein reduction on a 2-torus of
more or less any field theory that includes gravity.  Thus we see that
we actually learn something about ten-dimensional IIA supergravity by
compactifying it on a circle, namely that its global symmetries are
strongly suggestive of an eleven-dimensional origin.\footnote{There
are, of course, other more direct ways of seeing the
eleven-dimensional origin of IIA supergravity, but we are interested
here in considering the problem from the point of view of what can be
learned by toroidal compactification.}  Further toroidal
compactification to $D\le 8$ dimensions, however, seems to provide us
with lesser further insights into the properties of IIA supergravity
itself; the successive enlargements of the global symmetry groups as
one descends through the dimensions seem to be telling us more about
what happens under compactification than about the ten-dimensional
theory in its own right.

     In the case of the full ten-dimensional type IIA string theory,
the above discussion becomes, of course, more complicated.  In
essence, the main additional ingredient in the argument presented in
\cite{ht} is that there exist BPS-saturated states which can be
expected to be protected from quantum corrections by their
supersymmetry.  They correspond to $p$-brane soliton solutions in the
lower-dimensional supergravities that preserve half the supersymmetry.
Necessarily, these solutions form multiplets under the Cremmer-Julia
$E_{n(n)}$ symmetry, and after taking the analogue of the Dirac
quantisation condition between electric $p$-branes and magnetic
$(D-p-4)$-branes into account, one can argue that the appropriate
$E_{n(n)}(Z)$ discrete subgroup of the Cremmer-Julia group should also
be preserved in the full quantum theory.  The electrically-charged
NS-NS $p$-branes are identified with elementary states in the
perturbative string spectrum, while the magnetically-charged ones and
those with R-R charges will be non-perturbative states.  Thus one
conjectures an exact non-perturbative U-duality symmetry group
$E_{n(n)}(Z)$ \cite{ht}.  This suggests that the full
toroidally-compactified IIA string theory in $D\le 9$ has an
eleven-dimensional origin.

     As in our previous illustrative discussion of the classical
low-energy theories, it is not clear that the argument above acquires
significantly greater strength by considering toroidal compactifications to
$D\le8$ rather than simply the compactification to $D=9$ on a circle.
Further toroidal compactifications below $D=9$ introduce the additional
complexities of lower-dimensional theories while providing lesser further
insights into the structure of the IIA string itself.  In
fact, one might argue that some of the issues arising from the complexities
of the lower-dimensional theories may be serving to obscure rather than
clarify the relevant discussion. One aspect of this can be seen by
considering the Cremmer-Julia global symmetry groups for maximal
supergravities in $D\le 9$ dimensions.  Although these are normally
said to be $E_{n(n)}$ in $D=11-n$, it should be borne in mind that in
$D\le 7$ it is only true after dualising certain of the antisymmetric
tensor fields.\footnote{For $n\le 6$, the $E_{n}$ groups in their
non-compact versions have the following isomorphisms: $E_{5}\sim
SO(5,5)$, $E_4\sim SL(5,R)$, $E_3\sim SL(3,R)\times SL(2,R)$ and
$E_2\sim GL(2,R)$.}  To be precise, $E_{n(n)}$ is the symmetry group
if every field strength in $D\le 7$ is dualised whenever this results
in a field strength of a smaller degree.  For example, the version of
$D=7$ supergravity that has an $E_{4(4)}\sim SL(5,R)$ symmetry is the
one that comes from the dimensional reduction of $D=11$ supergravity
(or type IIA supergravity) followed by a dualisation of the 4-form
field strength $F_4$ in $D=7$ to give a 3-form field strength $F_3=
*F_4$. Strictly speaking, this is a different theory from the one in
which the 4-form is not dualised.  In particular, the potential $A_3$
for $F_4$ in this original version is non-locally related to the
potential $B_2$ for $F_3= *F_4$ in the dualised version.  Similar
considerations apply in lower dimensions too, with a rapidly-growing
variety of versions of the supergravity theories, whose potentials are
non-locally related.

     For many purposes the different versions of maximal supergravity in a
given dimension can be viewed as being essentially equivalent, if one
does not wish to relate the potentials in one version to those in
another.  The situation is a little different in string theory,
however, since at least at the perturbative level a particular
significance is attached to the 2-form potential $A_{\sst{ MN}}$ and
the metric tensor $g_{\sst{MN}}$ in ten dimensions, namely that these
NS-NS fields themselves are the ones that couple to the worldsheet of
the string.  By contrast, the $A_{\sst M}$ and $A_{\sst{MNP}}$
potentials, which are R-R fields, couple to the string only {\it via}
their field strengths.  Thus when discussing the T-duality symmetry of
the toroidally-compactified IIA string, it is important that whatever
other fields may require dualisations in order to implement the
symmetry, the field strengths associated with the NS-NS fields
$A_{\sst{MN}}$ and $g_{\sst{MN}}$ should be left intact and
undualised.  (Related issues in the context of the four-dimensional 
heterotic string have been considered in \cite{as}.)  It might further be 
argued that in all the discussions of
duality symmetries of the toroidally-compactified IIA string, one
should not require any of the field strengths originating from the
2-form potential or the metric in $D=10$ to be dualised.  In order to
explore this further, we shall consider the global symmetries
and the T-duality of the toroidal compactifications in detail.

     It is convenient to describe the toroidally-compactified type IIA
supergravities in a notation derived from the eleven-dimensional
supergravity that can be viewed as their progenitor.  Thus starting from
the metric $g_{\sst{MN}}$ and 3-form potential $A_{\sst{MNP}}$ in $D=11$,
we may consider a set of successive 1-step reductions on circles.  In each
reduction step from $(D+1)$ to $D$ dimensions, the metric in $(D+1)$ will
give rise to a metric, a Kaluza-Klein vector potential ${\cal A}_{\sst
M}$, and a ``dilatonic'' scalar field $\phi$ in $D$ dimensions.  An
$n$-index gauge potential in $(D+1)$ dimensions will give rise to an
$n$-index gauge potential and an $(n-1)$-index gauge potential in $D$
dimensions.  Thus it is easy to see that eventually after descending from
$D=11$ to $D$ dimensions, we shall have the following structure of
$D$-dimensional bosonic fields:
\bea
g_{\sst{MN}} &\longrightarrow & g_{\sst{MN}}\ ,\qquad \vec\phi\ ,\qquad 
{\cal A}_1^{(i)}\ , \qquad {\cal A}_0^{(ij)} \ ,\nn\\
A_3 &\longrightarrow & A_3\ ,\qquad A_2^{(i)}\ , \qquad A_1^{(ij)}\ ,
\qquad A_0^{(ijk)}\ ,\label{dfields}
\eea
where the indices $i, j, k$ run over the $11-D$ internal
toroidally-compactified dimensions, starting from $i=1$ for the step
from $D=11$ to $D=10$.  The subscripts on gauge potentials denote the
number of spacetime indices they carry.  The potentials $A_1^{(ij)}$
and $A_0^{(ijk)}$ are automatically antisymmetric in their internal
indices, whereas the 0-form potentials ${\cal A}_0^{(ij)}$ that come
from the subsequent dimensional reductions of the Kaluza-Klein vector
potentials ${\cal A}_1^{(i)}$ are defined only for $j>i$.  The
quantity $\vec\phi$ denotes the $(11-D)$-vector of dilatonic scalar
fields coming from the diagonal components of the internal metric.

     The Lagrangian for the bosonic $D$-dimensional toroidal
compactification of eleven-dimensional supergravity then takes the form
\cite{lpsol}
\bea
{\cal L} &=& eR -\ft12 e\, (\del\vec\phi)^2 -\ft1{48}e\, e^{\vec a\cdot 
\vec\phi}\, F_4^2 -\ft{1}{12} e\sum_i 
e^{\vec a_i\cdot \vec\phi}\, (F_3^{i})^2
-\ft14 e\, \sum_{i<j} e^{\vec a_{ij}\cdot \vec\phi}\, (F_2^{ij})^2
\nonumber\\
&& -\ft14e\, \sum_i e^{\vec b_i\cdot \vec\phi}\, ({\cal F}_2^i)^2
-\ft12 e\, \sum_{i<j<k} e^{\vec a_{ijk} \cdot\vec \phi}\,
(F_1^{ijk})^2 -\ft12e\, \sum_{i<j} e^{\vec b_{ij}\cdot \vec\phi}\,
({\cal F}_1^{ij})^2 + {\cal L}_{\sst{FFA}}\ ,\label{dgenlag}
\eea
where the ``dilaton vectors'' $\vec a$, $\vec a_i$, $\vec a_{ij}$, 
$\vec a_{ijk}$,
$\vec b_i$, $\vec b_{ij}$ are constants that characterise the couplings of
the dilatonic scalars $\vec \phi$ to the various gauge fields.  The
field strengths are associated with the gauge potentials in the obvious
way; for example $F_4$ is the field strength for $A_3$, $F_3^{(i)}$ is
the field strength for $A_2^{(i)}$, {\it etc}.  In general, the field
strengths appearing in the kinetic terms are not simply the exterior
derivatives of their associated potentials, but have Chern-Simons type
corrections as well.  On the other hand the terms included in ${\cal
L}_{\sst{FFA}}$, which denotes the dimensional reduction of the
$F_4\wedge F_4\wedge A_3$ term in $D=11$, are expressed purely in terms
of the potentials and their exterior derivatives.  The complete details of
all the terms, and the dilaton vectors, are given in \cite{lpsol}.

     It is important for our purposes to distinguish between the NS-NS and
the R-R fields in the lower-dimensional theories, viewed as
compactifications of the type IIA string.  In $D=10$, in the above
notation, we have NS-NS fields $g_{\sst{MN}}$, $\phi$ and $A_2^{(1)}$, and
R-R fields $A_3$ and ${\cal A}_1^{(1)}$.  Upon reduction to $D\le 9$, each
of these fields gives rise to sets of fields that have the same NS-NS or
R-R characteristics as their $D=10$ progenitors.  Thus if we split the
$i$, $j,\ldots$ indices as $i=(1,\a)$, {\it etc}, where $\a$,
$\beta,\ldots$ run from 2 to $11-D$, we have the following assignments of
NS-NS and R-R fields in $D$ dimensions \cite{lpsweyl}:
\bea
\hbox{ NS-NS}: && g_{\sst{MN}}\ ,\qquad \vec\phi\ ,\qquad 
A_2^{(1)}\ ,\qquad
A_1^{(1\a)}\ ,\qquad A_0^{(1\a\b)}\ ,\qquad {\cal A}_1^{(\a)}\ ,
\qquad {\cal A}_0^{(\a\b)}\nn\\
\hbox{R-R}: && A_3\ ,\qquad A_2^{(\a)}\ ,\qquad A_1^{(\a\b)}\ ,\qquad
A_0^{(\a\b\gamma)}\ ,\qquad {\cal A}_1^{(1)}\ , \qquad {\cal A}_0^{(1\a)}
\label{nsr}
\eea
The multiplicities of gauge potentials with one, two or three $\a$-type
indices are clearly $10-D$, $\ft12(10-D)(9-D)$ or $\ft16(10-D)(9-D)(8-D)$
respectively.

     It is known that in the customary dualised formulations of the
toroidally-compactified theories, where field strengths of higher rank are
dualised to ones of lower rank wherever possible, there is a T-duality
symmetry $D_{10-\sst D}\sim O(10-D,10-D)$, which is valid order by order 
in string perturbation theory \cite{bus}.  Since in low dimensions the
customary discussion includes a dualisation of the field strength
$F_3^{(1)}$ associated with the NS-NS 2-form potential $A_2^{(1)}$, which
in our discussion we wish to leave undualised, our first task will be to
verify that T-duality continues to operate in the same manner if we insist
that $F_3^{(1)}$ be left intact and undualised.  Likewise, we should
verify that none of the other NS-NS gauge fields have to undergo
dualisations in order to achieve the symmetry under T-duality
transformations.  In order to do this, it will be convenient to discuss
first a certain discrete subgroup of the putative T-duality group,
namely its Weyl group $W\!D_{10-D}$.  The Weyl groups of the T-duality
and U-duality groups for the toroidally-compactified strings were
discussed in detail in \cite{lpsweyl}.  They can be viewed as capturing the
essence of the full continuous symmetry groups of the supergravity
theories, and correspond precisely to the discrete subgroups that implement
exact permutations between the various field strengths, which, for more
general continuous group elements, would instead be mixed together in
more complicated ways.  (For example, in a case such as an
electric-magnetic duality in $D=4$, the Weyl group would be the
discrete $Z_2$ subgroup of duality rotations that implemented an exact
interchange of the electric and magnetic fields.)   If the theory is to
be invariant under such permutations of the field strengths, then, as can
be seen from (\ref{dgenlag}), the dilaton vectors associated with the
field strengths must permute at the same time.  Thus the Weyl group of
the T-duality group can be identified by seeking discrete sets of
permutations that exchange certain sets of dilaton vectors.

     From the details of the dilaton vectors given in \cite{lpsol}, it is
relatively straightforward to find their multiplet structures under the
relevant T-duality symmetries.  The procedure was described in detail in
\cite{lpsweyl}, and the only difference here is that we are paying close
attention to the question of which fields must be dualised in order to
achieve the $W\!D_{10-\sst D}$ symmetry, and which can be left undualised.
To begin, we tabulate the multiplicities of the various gauge potentials
appearing in the toroidally-compactified theories:

\bigskip\bigskip

\centerline{
\begin{tabular}{|c|c|c|c|c|c|c|c|c|c|c|c|}\hline
$D$ &  \multicolumn{5}{c|}{NS-NS} & 
\multicolumn{6}{c|}{R-R} \\ \hline
 & $A_2^{(1)}$ & $A_1^{(1\a)}$ & ${\cal A}_1^{(\a)}$ & $A_0^{(1\a\b)}$ &
${\cal A}_0^{(\a\b)}$ & $A_3$ & $A_2^{(\a)}$ & $A_1^{(\a\b)}$ &
${\cal A}_1^{(1)}$ & $A_0^{(\a\b\gamma)}$ & ${\cal A}_0^{(1\a)}$
\\ \hline\hline 
8 & 1 & 2 & 2 & 1 & 1 & 1 & 2 & 1 & 1 & -- & 2 \\ \hline
7 & 1 & 3 & 3 & 3 & 3 & 1 & 3 & 3 & 1 & 1 & 3 \\ \hline
6 & 1 & 4 & 4 & 6 & 6 & 1 & 4 & 6 & 1 & 4 & 4 \\ \hline
5 & 1 & 5 & 5 & 10 & 10 & 1 & 5 & 10 & 1 & 10 & 5 \\ \hline
4 & 1 & 6 & 6 & 15 & 15 & 1 & 6 & 15 & 1 & 20 & 6 \\ \hline
3 & 1 & 7 & 7 & 21 & 21 & -- & 7 & 21 & 1 & 35 & 7 \\ \hline
\end{tabular}}
\bigskip

\centerline{Table 2: Multiplicities of gauge potentials in compactified 
IIA strings}
\bigskip\bigskip
     
     At this stage, the discussion separates into two parts, one for the
NS-NS potentials and the other for the R-R potentials.  The situation is
simpler for the NS-NS potentials since, as remarked previously, it turns
out that no dualisations are necessary in order to assemble these fields
into multiplets under T-duality.  Thus we find that the multiplets under
the $W\!D_{10-\sst D}$ Weyl group for the NS-NS potentials are as follows:

\bigskip\bigskip

\centerline{
\begin{tabular}{|c|c|c|c|c|}\hline
$D$ &  Weyl Gp & $ A_2^{(1)}$ & 
$\{ A_1^{(1\a)}, {\cal A}_1^{(\a)} \}$ & 
$\{ A_0^{(1\a\b)}, {\cal A}_0^{(\a\b)} \}$ \\ \hline
8 & $W\!D_2$ & (1,1) & (2,2) & (1,2) + (2,1) \\ \hline
7 & $W\!D_3$ & 1 & 6 & 12 \\ \hline
6 & $W\!D_4$ & 1 + 1 & 8 & 24 \\ \hline
5 & $W\!D_5$ & 1 & 10 & 40 \\ \hline
4 & $W\!D_6$ & 1 & 12 + 12 & 60 \\ \hline
3 & $W\!D_7$ & 1 & 14 & 84 \\ \hline
\end{tabular}}
\bigskip

\centerline{Table 3: T-duality Weyl-group multiplets for NS-NS potentials
in IIA strings}
\bigskip\bigskip

Some comments about the results in this table are in order.  First, we
note that the T-duality group in $D=8$ dimensions is $D_2\sim O(2,2)$,
which is a product group, $D_2\sim D_1\times D_1 \sim O(2,1) \times
O(2,1)$.  Consequently, the multiplicities in $D=8$ are given in terms
of their dimensions under the two factors.  Secondly, as discussed in
\cite{lpsweyl}, the multiplicities for 0-form potentials acquire a
doubling when the Weyl-group multiplets are assembled.  This is
because there is a discrete symmetry $\vec\phi\rightarrow
-\vec\phi$ and $A_0\rightarrow e^{\vec c\cdot\vec\phi}\, A_0$, where
$\vec c$ denotes the dilaton vector associated with the 0-form
potential $A_0$.  Thus, for example, although we see from Table 2 that
there are three NS-NS 0-forms $A_0^{(1\a\b)}$ and three NS-NS 0-forms
${\cal A}_0^{(\a\b)}$ in $D=7$, the permutations of their dilaton vectors
under the $W\!D_3$ Weyl group form an irreducible 12-element
representation, involving the six dilaton vectors and their negatives. 
Thirdly, a doubling of multiplicities occurs for a different reason in the
case of the 2-form potential $A_2^{(1)}$ in $D=6$ and the 1-form potentials
in $D=4$.  These doublings really only occur at the level of
solutions, and arise because in these cases the field strengths can
carry either electric or magnetic charges. Note however that in each
dimension, including $D=6$, the 2-form potential $A_2^{(1)}$ itself is
a singlet under the Weyl group.

     The discussion of the T-duality for the R-R potentials is more
complicated, because here we find that certain dualisations of field
strengths can become necessary in dimensions $D\le 8$, in order to
assemble the Weyl-group multiplets.  Since this is a dimension-dependent
procedure, we shall adopt a different format for presenting the results
here.  Grouping together R-R potentials into multiplets under the
T-duality Weyl group, we find the following:

\bigskip
\vfill\eject

\bea
D=8: && \{ A_3 \}_{(1,2)} \qquad \{ A_2^{(\a)} \}_{(2,1)} \qquad
\{ A_1^{(\a\b)} \}_{(1,1)} \qquad \{ {\cal A}_1^{(1)} \}_{(1,1)} 
\qquad \{ {\cal A}_0^{(1\a)} \}_{(1,2)+(2,1)} \nn\\
D=7: && \{ A_3,A_2^{(\a)} \}_4  \qquad \{A_1^{(\a\b)}, {\cal A}_1^{(1)}
\}_4 \qquad \{ A_0^{(\a\b\gamma)}, {\cal A}_0^{(1\a)} \}_{4+4} \nn\\
D=6: && \{A_3,A_1^{(\a\b)},{\cal A}_1^{(1)} \}_8 \qquad \{ A_2^{(\a)} \}_8
\qquad \{ A_0^{(\a\b\gamma)}, {\cal A}_0^{(1\a)} \}_{8+8}\nn\\
D=5: && \{A_3, A_0^{(\a\b\gamma)}, {\cal A}_0^{(1\a)} \}_{16+16}\qquad
\{ A_2^{(\a)}, A_1^{(\a\b)},{\cal A}_1^{(1)} \}_{16}\nn\\
D=4: && \{ A_3 \}_1 \qquad 
\{ A_2^{(\a)},A_0^{(\a\b\gamma)},{\cal A}_0^{(1\a)} \}_{32+32} \qquad
\{ A_1^{(\a\b)},{\cal A}_1^{(1)} \}_{32}\nn\\
D=3: && \{ A_2^{(\a)} \}_{14} \qquad 
\{ A_1^{(\a\b)},{\cal A}_1^{(1)},
A_0^{(\a\b\gamma)}, {\cal A}_0^{(1\a)} \}_{64+64}\nn
\eea
\centerline{Table 4: T-duality Weyl-group multiplets for R-R potentials
in  IIA strings}
\bigskip

The subscripts denote the dimensions of the representations under
the relevant T-duality Weyl groups given previously.  In cases where gauge
potentials of different degrees are grouped together within a set of
braces, this implies that dualisations must be performed.  

     To summarise the story so far, we have seen that the Weyl groups of
the T-duality symmetries of toroidally-compactified type IIA strings, which
are $W\!D_{10-\sst D}$ in the usual dualised versions with $E_{11-\sst
D}$ U-duality, continue to be the same if we work instead with the the
versions of the theories in which no NS-NS fields are dualised.\footnote{By
arguments analogous to those discussed in \cite{lpsweyl}, this will
continue to be true for the full T-duality groups themselves.}  This is
an important point since T-dualities are perturbative symmetries, and at
least at the perturbative level, the bare NS-NS fields $g_{\sst{MN}}$ and
$A_2^{(1)}$ have intrinsic significance as the fields that couple
directly to the string worldsheet.  By contrast, obtaining the
$W\!D_{10-\sst D}$ symmetry does require that some of the R-R field
strengths be dualised; this does not present any difficulty since the
associated R-R-potentials couple to the string only {\it via} their field
strengths, and thus the theory could equally well be formulated in such a
way that it is the relevant dualised field strengths that couple to the
string. 

     Having discussed the T-dualities of the compactified theories, we now
turn to a consideration of their U-dualities.  As we remarked previously,
there are many different versions of $D$-dimensional maximal
supergravity, corresponding to different choices of duality
complexions for the various field strengths.  At the level of the
gauge potentials, which are the basic field variables in the actions,
these different versions are related to one another by non-local
field redefinitions.  The global symmetry groups for the
different versions will be different too, and the conventional $E_{11-\sst
D}$ Cremmer-Julia symmetries are associated with the versions where field
strengths are always dualised if it results in a lowering of their degrees. 
Here, we shall pursue the investigation of what happens to the U-duality if
we again opt for versions of the compactified theories in which NS-NS fields
remain undualised.  

     Let us recall that the Cremmer-Julia symmetry can first of all be
thought of as the global symmetry group of the scalar sector of the
supergravity Lagrangian, \ie the part that involves just the dilatonic
scalars $\vec\phi$ and the axionic scalars $A_0^{(ijk)}$ and ${\cal
A}_0^{(ij)}$.  The invariance of the entire theory under this symmetry can
then be understood as the consequence of a covariance of the Lagrangian,
with the scalars coupling to the other fields {\via} the metric or vielbeins
on the scalar manifold. For example, in $D=9$ the scalar sector of the
Lagrangian takes the form 
\be
e^{-1}\, {\cal L} = -\ft12(\del\varphi)^2 -\ft12 (\del \phi)^2 -\ft12
e^{-2\phi}\, (\del \chi)^2\ ,\label{d9scal}
\ee
where in our type IIA notation $\chi$ is the axion ${\cal A}_0^{(1,2)}$,
and $\varphi$ and $\phi$ are related to $\phi_1$ and $\phi_2$ by
$\varphi=-\ft14\sqrt7\phi_1-\ft34\phi_2$ and $\phi=\ft34\phi_1 -\ft14\sqrt7
\phi_2$. This has a $GL(2,R)\sim SL(2,R)\times R$ global symmetry, where
$SL(2,R)$ acts on $\tau=\chi + i e^\phi$ by fractional linear
transformations, and the additional $R$ symmetry comes from constant
shifts of $\varphi$.  Thus included in this $GL(2,R)$ Cremmer-Julia
symmetry is the ``gauge transformation'' $\chi\rightarrow \chi\, +$ const.
for the 0-form potential (corresponding to $\tau\rightarrow\tau\, +$ const.
in $SL(2,R)$).  Unlike a gauge transformation for a higher-rank potential,
this gauge transformation is ``demoted'' to a global symmetry, since the only
closed 0-form is a constant.  The purpose of making this observation is
to emphasise that when we look for global symmetries in lower
dimensions, we should recognise that such global shift symmetries of 0-form
potentials should be included in the symmetry group, whereas the local
gauge transformations of higher-rank potentials should not.

     First, we shall consider the global symmetries for the versions
of the toroidally compactified $D=11$ theory in which no dualisations
at all are performed.  It is not hard to see that in $D$ dimensions
there will be an ``obvious'' $GL(11-D,R)$ symmetry, generalising the
$GL(2,R)$ symmetry in $D=9$.  There will however be more than this,
since as we descend through the dimensions we accumulate more 0-form
potentials than are needed for the invariances of the
$GL(11-D,R)$-symmetric scalar manifold.  The excess 0-forms, with
global shift symmetries, will contribute additional $R$ factors in the
global symmetry group of the entire scalar manifold.  In fact the
excess 0-forms are precisely the axions $A_0^{(ijk)}$ coming from the
dimensional reduction of $A_3$ in $D=11$, of which there will be
$p=(1, 4, 10, 20, 35, 56)$ in $D=(8, 7, 6, 5, 4, 3)$ respectively.
Thus in their totally undualised formulations, the maximal
supergravities in $D$ dimensions have global $R^p* GL(11-D,R)$.  Here $*$
denotes a semi-direct product, which arises rather than a direct product
because the scalar fields $A_0^{(ijk)}$ carry $GL(11-D, R)$ indices,
and thus transform (linearly) under $GL(11-D, R)$.

     In the above discussion, all the axionic fields $A_0^{(ijk)}$ are
placed on an equal footing.  Not surprisingly, this runs into
difficulties with the T-duality that we discussed earlier, since we
were required to put together the R-R axions $A_0^{(\a\b\gamma)}$ and
${\cal A}_0^{(1\a)}$ in order to form representations under the
$O(10-D,10-D)$ T-duality group.  However, it can be shown that
although it is indeed possible to make a choice of field variables in
which all the $A_0^{(ijk)}$ axions are simultaneously covered
everywhere by derivatives, and are thus subject to manifest shift
symmetries, the price that is paid is that no other axions can also at
the same time be covered with derivatives everywhere \cite{clpst}.  In
particular, in the totally undualised formulation the R-R axions
$A_0^{(\a\b\gamma)}$ can be covered by derivatives everywhere but then
their would-be T-duality multiplet partners ${\cal A}_0^{(1\a)}$
cannot be also covered everywhere. This would break the T-duality
symmetry.\footnote{T-duality implies that each of the R-R scalars
$A_0^{(\alpha\beta\gamma)}$ and ${\cal A}_0^{1\alpha}$ has a shift R
symmetry.  Since these R symmetries are commuting, it follows that
there must exist a choice of field variables such that these R
symmetries are manifest, {\it i.e.}\ all the R-R scalars are
similtaneously covered by derivatives.  However, this is not possible
in the undualised form of supergravities, and hence the T-duality is
broken.  On the other hand, in the R-R dualised form, it is possible
to redefine field variables such that all the R-R fields are covered
by exterior derivatives.  Thus it is worth emphasising that
field redefinitions can make the global symmetries manifest, but do
not alter whether or not they are present.  It is the dualisation
procedure that alters the global symmetries in the different versions
of the supergravities.}  Another way of seeing this is to note that the
$GL(11-D,R)$ symmetry of the totally undualised formulation mixes the
$i=1$ index value with the $i=\a$ values, and thus it interchanges
NS-NS fields with R-R fields.  The $GL(10-D,R)$ subgroup that acts only on
the $i=\a$ internal indices, and thus preserves the NS-NS and R-R
sectors independently, is not large enough to contain the
$O(10-D,10-D)$ T-duality group.

     In order to try to find an extended symmetry group, possibly with the
inclusion of non-perturbative generators, that includes the T-duality
group, we may instead opt for a choice of field variables in which all
the R-R potentials are simultaneously covered by derivatives.  In fact
it was shown in \cite{clpst} that there always exists such a choice of
field variables, when necessary R-R fields are dualised.  This means
that we can still dualise the R-R fields as required for the T-duality
discussion, and at the same time have all the R-R axions subject to
shift symmetries.  Note that T-duality, as a perturbative symmetry, is
usually described in the context of heterotic string theory, or the
NS-NS sector of type II strings, where there are no R-R fields and so
the symmetry is concerned purely with perturbative degrees of freedom.
In the absence of R-R fields, there is a maximal $R^{(10-D)(9-D)/2}$ symmetry
contained within the T-duality group $O(10-D, 10-D)$, in that there
are a total of $(10-D)(9-D)/2$ NS-NS scalars appearing in the 
Lagrangian only {\it via}
derivatives.  In type II theories the introduction of R-R fields, 
which are believed to be still organised by T-duality despite their
non-perturbative nature, has a subtle effect in modifying the
T-duality group.  It was shown in \cite{clpst} that although all the
R-R potentials can be covered by derivatives, this is at the price
that none of the NS-NS scalars appears in the full Lagrangian only
through a derivative any more.  Thus the previously manifest
$R^{(10-D)(9-D)/2}$ symmetry in the T-duality group $O(10-D, 10-d)$ of the
heterotic string becomes no longer manifest with the introduction of
the R-R fields in the type IIA theory.  Of course, the R symmetries in
$O(10-D, 10-D)$ are actually still preserved, although non-manifestly,
even in the presence of the R-R fields since the R-R axions, as in the case
of all higher-form gauge potentials, transform linearly under the T-duality
group $O(10-D, 10-D)$.  Since now there are $q=(2,4,8,16,32,64)$ R-R
axions in $D=(8,7,6,5,4,3)$ dimensions, which appear in the Lagrangian
through their 1-form field strengths only, we therefore have the
global symmetries $R^q*O(10-D,10-D)$ in these R-R
dualised versions of the lower-dimensional maximal supergravities.
Here $*$ denotes a semi-direct product.  The reason why it is a
semi-direct product rather than a direct product is that the R-R
axions rotate (linearly) among themselves under the T-duality $O(10-D,
10-D)$, in the same manner as higher-form gauge potentials, and thus
the $R^q$ symmetry does not commute with $O(10-D, 10-D)$.\footnote{We
are grateful to E. Cremmer for extensive discussions on the
semi-direct product structure of the symmetry groups.}

    As a consistency check, we may verify that with the global symmetry
groups taken to be $G=R^q * O(10-D,10-D)$, we indeed get the correct
counting of scalars if we augment them with extra scalars in the adjoint
representation of the maximal compact subgroup $H$ of $G$, such that the
physical scalars are in the coset $G/H$.  Thus we have
$H=O(10-D)\times O(10-D)$, implying that $G/H$ has dimension
$\{6,13,24,41,68,113\}$ in $D=\{8,7,6,5,4,3\}$.  This is exactly
coincident with the total numbers of scalars in each dimension, where
axions, dilatonic scalars and the dualisations of rank-$(D-1)$ R-R
field strengths (where appropriate) are included, but the dualisations
of any NS-NS field strengths are excluded.  (The numbers of axions and
R-R fields dual to axions can be read off from table 2.)  Furthermore,
the $D=10$ dilaton itself is left out of the counting, since it does
not participate in the perturbative symmetries described by the $R^q *
O(10-D,10-D)$ group.  The counting of the various scalars in the
cosets is summarised in table 5 below, where, in the columns headed by
NS-NS and R-R, the numbers of axions of each type are listed:

\bigskip\bigskip

\centerline{
\begin{tabular}{|c|c|c|c|c|c|}\hline
  &  \multicolumn{2}{c|}{ Coset} & 
\multicolumn{3}{c|}{Numbers of scalars} \\ \hline
$D$ & $G/H$ & Dimension & NS-NS & R-R &
Dilatons\\ \hline\hline 
8 & $R^2 * O(2,2)/O(2)\times O(2)$ & 6 & 2 & 2 & 2 \\ \hline
7 & $R^4 * O(3,3)/O(3)\times O(3)$ & 13 & 6 & 4 & 3 \\ \hline
6 & $R^8 * O(4,4)/O(4)\times O(4)$ & 24 & 12 & 8 & 4 \\ \hline
5 & $R^{16} * O(5,5)/O(5)\times O(5)$ & 41 & 20 & 16 & 5 \\ \hline
4 & $R^{32} * O(6,6)/O(6)\times O(6)$ & 68 & 30 & 32 & 6 \\ \hline
3 & $R^{64} * O(7,7)/O(7)\times O(7)$ & 113 & 42 & 64 & 7 \\ \hline
\end{tabular}}
\bigskip

\centerline{Table 5: Counting of scalars in $G/H$ for R-R dualisation}
\bigskip\bigskip

     To recapitulate the situation so far, we have seen that by working in
the formulations of the supergravity theories where all the R-R fields are 
covered with derivatives, we have perturbative symmetry groups $R^q * 
O(10-D,10-D)$ in each dimension.  We may now discuss the possible 
enlargements of these symmetry groups to include non-perturbative 
generators as well.  However, in keeping with the spirit of our previous 
discussions, we shall require that these non-perturbative symmetries be 
achieved by dualising only R-R fields, but not NS-NS fields.  The results 
for these non-perturbative global symmetry groups may be presented in the 
following table, below which we shall discuss the various entries:

\bigskip\bigskip

\centerline{
\begin{tabular}{|c|c|c|c|c|}\hline
 &  \multicolumn{3}{c|}{Global Symmetry Groups} & T-duality\\ \hline
$D$& No dualisation & R-R dualisation &Full dualisation & \\ \hline\hline
9 & $GL(2,R)$ & $GL(2,R)$ & $GL(2,R)$ & --\\ \hline
8 & $R*GL(3,R)$ &$SL(3,R)\times SL(2,R)$ & $SL(3,R)\times SL(2,R)$ 
& $O(2,2)$\\ \hline 
7 & $R^4*GL(4,R)$ & SL(5,R) & $SL(5,R)$ & $O(3,3)$\\ \hline 
6 & $R^{10} * GL(5,R)$ & $R^8 * O(4,4)$ &
$SO(5,5)$ & $O(4,4)$\\ \hline 
5 & $R^{20} * GL(6,R)$ & $R^{16} * O(5,5)$ & 
$E_{6(6)}$ & $O(5,5)$\\ \hline 
4 & $R^{35} * GL(7,R)$ &
$R^{32} * O(6,6)$ & $E_{7(7)}$ & $O(6,6)$\\ \hline 
3 & $R^{56} * GL(8,R)$ & $R^{64} *  O(7,7)$ &
$E_{8(8)}$ & $O(7,7)$\\ \hline
\end{tabular}}
\bigskip

\centerline{Table 6: Global symmetry groups for maximal supergravities}
\bigskip\bigskip

\noindent No dualisation is necessary in $D=9$, whilst for $D=8$ and 7
the only field that needs to be dualised is the R-R 4-form field
strength. Thus for $D\ge 7$, the full non-perturbative Cremmer-Julia
group $E_{11-D}$ is consistent with the requirement that no NS-NS
fields should be dualised.  This implies in particular that $R^q *
O(10-D,10-D)$ is a perturbative subgroup of the non-perturbative
$E_{11-D}$ group in these cases.  For $D\le 6$, the story becomes more
complicated, in that the existence of the $E_{11-D}$ symmetry requires
the dualisation of NS-NS fields as well as R-R fields.  If we instead
insist that only R-R fields can be dualised, which is consistent with
T-duality, we find that the symmetry group is simply $R^q * O(10-D,
10-D)$, which is perturbative only.  Adding any non-perturbative
generators, such as rotations between NS-NS and R-R fields, would
naturally force the NS-NS fields to be dualised, since the R-R fields
form multiplets together with their duals.  Thus it seems that for
$D\le 6$, the requirement that no NS-NS fields be dualised rules out
the possibility of having any enlargement of the perturbative $R^q *
O(10-D,10-D)$ symmetry to include non-perturbative generators.  In
fact, the only way to find a non-perturbative symmetry, while leaving
the NS-NS fields undualised, is in the case where no fields at all
undergo dualisations.  (We shall show presently that this is
consistent with a possible eleven-dimensional supermembrane \cite{bst}
origin.)  The $GL(11-D, R)$ and $O(10-D, 10-D)$ groups are subgroups
of $E_{11-D}$, and indeed the closure of the two groups generates the
full $E_{11-D}$ group.  However, the $R^p$ and $R^q$ symmetries are
not contained within $E_{11-D}$ in lower dimensions.  In fact the maximal
numbers of shift symmetries in the $E_{11-D}$-symmetric versions of the
supergravity theories have been studied in \cite{frecg}.  These numbers in
$D=\{8,7,6,5,4,3\}$ are $\{3,6,10,16,27,44\}$, of which $2^{8-D} = \ft12 q$
are for R-R axions, with the remainder being for NS-NS axions.  In
particular, it is evident that in the lower dimensions these numbers are
less than the numbers $p$ or $q$ of shift symmetries in the undualised or
R-R dualised formulations, demonstrating that the symmetry groups in these
versions of the supergravities cannot be subgroups of the
symmetries of the $E_{11-D}$-symmetric versions.
 
    Note that a similar consistency check to that which we performed for
the supergravities with R-R dualisation can be carried out also for the
$G=R^p * GL(11-D,R)$ global symmetries of the totally undualised supergravity
theories.  In this case the corresponding maximal compact subgroups of $G$
are given by $H=O(11-D,R)$, and so the coset $G/H$ has dimension
$\{3,7,14,25,41,63,92\}$ in $D=\{8,7,6,5,4,3\}$.  From the results in table
2, we see that these dimensions are indeed equal to the numbers of axions
plus dilatons in each dimension, where in this case no dualisations at all
of $(D-1)$-form field strengths are performed.  Note also that in this
case, since the $G=R^p * GL(11-D,R)$ symmetry is non-perturbative, the
$D=10$ dilaton itself is included in the counting of dilatons.  The
counting of the various scalars in the cosets is presented in table 7
below:

\bigskip\bigskip

\centerline{
\begin{tabular}{|c|c|c|c|c|c|}\hline
  &  \multicolumn{2}{c|}{ Coset} & 
\multicolumn{3}{c|}{Numbers of scalars} \\ \hline
$D$ & $G/H$ & Dimension & NS-NS & R-R &
Dilatons\\ \hline\hline 
8 & $R * GL(3,R)/O(3)$ & 7 & 2 & 2 & 3 \\ \hline
7 & $R^4 * GL(4,R)/O(4)$ & 14 & 6 & 4 & 4 \\ \hline
6 & $R^{10} * GL(5,R)/O(5)$ & 25 & 12 & 8 & 5 \\ \hline
5 & $R^{20} * GL(6,R)/O(6)$ & 41 & 20 & 15 & 6 \\ \hline
4 & $R^{35} * GL(7,R)/O(7)$ & 63 & 30 & 26 & 7 \\ \hline
3 & $R^{56} * GL(8,R)/O(8)$ & 92 & 42 & 42 & 8 \\ \hline
\end{tabular}}
\bigskip

\centerline{Table 7: Counting of scalars in $G/H$ for no dualisation}
\bigskip\bigskip

     Thus in summary, if we insist that no dualisations of NS-NS fields be 
performed, we can follow two possible routes.  In one of them, we look for 
the largest symmetry group that includes the perturbative T-duality symmetry 
$O(10-D,10-D)$.  We find that this is the perturbative symmetry group $R^q * 
O(10-D,10-D)$, where $q=2^{9-D}$ is the number of R-R axions, for general 
values of $D$.  In $D\ge 7$ it can be enlarged to include non-perturbative 
generators, since these still do not require the dualisation of NS-NS 
fields.  Below seven dimensions, no enlargement to such a non-perturbative 
group is possible.  The results are summarised in the ``R-R dualisation'' 
column in table 6.  A second, alternative, route is to sacrifice the 
T-duality symmetry as a subgroup, in order to find a non-perturbative
symmetry group in lower dimensions.  Indeed we can then find
non-perturbative symmetries that exist also in $D\le 6$, by choosing the
versions of the supergravity theories in which no dualisations at all are
performed. The results in this case are summarised in the ``No 
dualisation" column in table 6.  In $D\le 8$, this is achieved at the price 
of no longer having the perturbative $O(10-D, 10-D)$ T-duality symmetry as a
subgroup.

     It should be emphasised that in the versions of the $D$-dimensional
supergravities where no dualisations at all are performed, the
$R^p*GL(11-D,R)$ global groups are symmetries of the Lagrangian for all
values of $D$.  On the other hand, in the R-R dualised versions of the
supergravities, the $R^q*O(10-D,10-D)$ groups are symmetries of the
Lagrangian itself only when $D$ is odd.  When $D$ is even, the symmetries
are valid only at the level of the equations of motion.  This is similar to
the situation for the $E_{11-D}$-symmetric versions of the supergravities.

     A general remark about the nature of the various symmetry groups is
perhaps in order here.  One might have thought that at the level of the
equations of motion all the potentials would be occurring {\it via} their
field strengths, and that therefore there should be a unique answer for
the global symmetry group of the theory.  As we have seen, this is in fact
not the case.  The explanation lies in the fact that as a result of the
Chern-Simons modifications to the field strengths, bare potentials that
are not covered by derivatives appear even in the field equations.  By
performing field redefinitions, it is possible to ensure that all the
fields that need to be dualised in the usual $E_{11-\sst D}$-symmetric
formulations of the supergravity theories are everywhere covered with
derivatives.  However, if instead some of these fields are ``sacrificed,''
it is possible to perform field redefinitions that transfer the
derivatives onto other potentials, thereby gaining new symmetries at the
cost of losing the previous ones.  (Once field redefinitions are performed
that cause potentials to be exposed without derivatives, symmetries that
require dualisations of the associated field strengths no longer act
locally on the fields of the theory.)  Thus for example in four dimensions
the usual fully-dualised theory has a global $E_7$ symmetry, whose maximal
abelian invariant subalgebra has dimension 27 (\ie shift symmetries for 16
R-R axions and 11 NS-NS axions) \cite{frecg}.  If the symmetries involving
dualisations of the NS-NS field strengths are sacrificed, the derivatives
on their potentials can be transferred to other fields, including all 32
R-R axions.  The $R^{32}*O(6,6)$ symmetry for this choice is manifestly
not a subgroup of $E_7$.  If all symmetries involving dualisations, both
for NS-NS and R-R fields are sacrificed, the derivatives can be
transferred so as to cover all of the 35 axions $A_0^{(ijk)}$, and the
resulting global symmetry group $R^{35}*GL(7,R)$ is neither contained in,
nor does it contain, either of the previous two groups.  Thus the global
symmetry group of the theory is not uniquely determined until the fields
on which it is to act locally are specified.  For example in four
dimensions, if it is to act only on potentials then the symmetry is
$R^{35}*GL(7,R)$; it is instead $R^{32}*O(6,6)$ if it is to act on NS-NS
potentials but R-R field strengths; and it is $E_7$ if it is to act on
field strengths of both types.  String theory, at least in a perturbative
formulation, favours the first or second possibilities.  Since
perturbative string theory the R-R fields coupiling only {\it via} their
field strengths, the second possibility seems to be the most natural one.

     One further remark that should be made is that the above discussions of 
the various possibilities for non-perturbative symmetry groups that do not 
require dualisations of NS-NS fields do not in any way conflict with the 
known multiplet structures of BPS-saturated soliton solutions.  In particular, 
the $p$-brane solutions that carry a single type of electric or magnetic
charge fall into multiplets under the Weyl group of the usual U-duality
group, an example being the black holes in $D=4$, which form
an irreducible 56-component representation under the Weyl group of $E_7$
\cite{lpsweyl}.  Under the Weyl group of $R^{32}*O(6,6)$ they will instead 
form a reducible $32+12+12$ component representation, while under 
$R^{35}*GL(7,R)$ they will form a reducible $7+7+21+21$ component 
representation.  In all of these cases, the total number of distinct black
hole species will be 56.  Similar considerations apply to all the other
species of $p$-brane solitons. 

     With the proposal that the degrees of freedom of string theory are
more appropriately described by an eleven-dimensional theory in
general, with the string providing a useful limiting description in
the perturbative regime, it is of interest to consider how the
discussions of T-duality and U-duality in this paper might apply in
M-theory itself.  It has sometimes been argued that the
eleven-dimensional supermembrane \cite{bst} should be viewed as a
fundamental entity in its own right, whose quantisation might give
rise to an eleven-dimensional theory which has $D=11$ supergravity as
its low-energy limit.  There are many objections to this viewpoint,
centering on the apparently insurmountable difficulties in setting up
a sensible perturbative quantisation scheme.  One signal of this
problem is the absence of any candidate for a loop-counting parameter,
analogous to the coupling constant $e^{-\phi_0}$ in string theory,
owing to the fact that there is no dilaton in the eleven-dimensional
theory.  If, as therefore seems likely, there is no perturbative
regime for the quantisation of a supermembrane, it is not clear that
there would be any utility to attempting a description of some region
of the modulus space of $M$-theory in terms of a fundamental
supermembrane.  Indeed, it can be argued that string theory is useful
only insofar as it enables a region of the modulus space of M-theory
to be described perturbatively \cite{w1}.

     Another indication of the intrinsically non-perturbative nature of
the supermembrane is provided by considering T-duality in terms of an
hypothetical $D=11$ supermembrane description.  We have seen above that
in the case of string theory, the usual T-duality groups of the
toroidally-compactified theory are perfectly compatible with a
formulation of the theory in which none of the NS-NS fields needs to be
dualised.  It is important that this should be possible because
T-duality is a perturbative symmetry, and thus should be adequately
describable in terms of string perturbation theory.  In this theory, the
NS-NS gauge potentials have a fundamental significance since it is they,
and not their field strengths, that couple to the worldsheet of the
string.  However, the situation would be quite different if the
lower-dimensional theories were to be described as toroidal
compactifications of a quantised supermembrane. The difference lies in
the fact that in such a supermembrane theory it is the 3-form potential
$A_3$ of $D=11$ supergravity, rather than the 2-form potential
$A_2^{(1)}$ of $D=10$ supergravity, that would now be playing a
distinguished r\^ole since $A_3$ couples directly to the world-volume
of the membrane.  But as we have seen, in order to realise the T-duality
symmetries it is necessary that some of the Kaluza-Klein descendants of
$F_4$, which from the string point of view are R-R fields, must be dualised. 
This means that T-duality must act non-locally on the
potential $A_3$ in eleven dimensions, and thus it would be incompatible
with a perturbative description in terms of fundamental membranes.

       The eleven-dimensional supermembrane action implies that the
symmetry group of the theory compactified on a torus should 
$R^p \times GL(11-D, R)$, obtained from the supergravity theories
where no dualisation is performed.  It is non-perturbative in nature.
However there is a subgroup of $GL(11-D, R)$, namely $O(10-D)$, which
rotates NS-NS and R-R field within themselves.  This $O(10-D)$ subgroup
is perturbative, since it is the intersection of $G(11-D, R)$ and 
$O(10-D,10-D)$.  This perturbative symmetry of the membrane action on a
torus may provide some clue as to how to quantise the theory, if it is
quantisable at all.

     In conclusion, we have seen that various different viewpoints are
possible concerning the global symmetry groups in toroidally-compactified
type IIA strings in lower dimensions.  In particular, when $D$ is less
than seven the symmetry groups obtained for the theories with only R-R
dualisations are different from, and are not contained in, the usual
$E_{n(n)}$ groups that arise for the theories where dualisations of NS-NS
fields are also performed.  At least at the level of the perturbative
T-duality symmetry it seems reasonable to avoid such NS-NS dualisations,
since it is the NS-NS potentials $A_{\sst{MN}}$ and $g_{\sst{MN}}$
themselves, rather than gauge-invariant field strengths built from them,
that couple directly to the string worldsheet.  If this reasoning is
extended to the discussion of non-perturbative symmetries too then it might
be argued that since the possible symmetry groups under the various
viewpoints are inequivalent, then it is not inconceivable that it is the 
symmetries in the formulations that do not require NS-NS dualisations that
should be regarded as more fundamental.\footnote{Another way to express the 
observation is that non-perturbative symmetries in
string theory are associated with non-local symmetries in the perturbative
description.  Thus for example the non-perturbative $SL(2,R)$ S-duality of
four-dimensional string theory, which preserves the NS-NS and R-R
sectors but is associated with dualisations of the 2-form field
strengths \cite{as}, acts non-locally on the NS-NS potentials.} Such
reasoning need not be in conflict with any of the results about the
relation between compactifications of M-theory and of the type IIA string,
since, as we observed in the introduction to the paper, the key results can
already be seen in the compactifications to nine dimensions.  Indeed, the
fact that the relations are robust with respect to the different choices
for the dualisations that should be performed in the lower-dimensional
theories serves to emphasise that the basic relations between the
higher-dimensional theories can be established without needing to invoke
the details of the lower-dimensional compactifications.  

     Finally, we remark that our discussions have concentrated on the
supergravity theories, rather than the full quantum string theories.  Thus
for example although we have used the language of string theory to
distinguish between symmetries that would act perturbatives, and those that
would act non-perturbatively, at the level of our discussion the symmetry
groups have all been the classical continuous groups of the Cremmer-Julia
kind.  It would be interesting to study the implications of the quantisation
conditions for electrically and magnetically charged solitons, to uncover
the likely discrete subgroups that would survive in the full string theory.

\section*{Acknowledgement}

     We are grateful to M.J. Duff, S. Mukherji and K.S. Stelle for 
discussions, and to E. Cremmer and 
B. Julia for extensive discussions on the global symmetry groups of 
different versions of the supergravity theories.


\end{document}